\input harvmac
\noblackbox
\def\ias{\vbox{\sl\centerline{School of Natural Sciences, 
Institute for Advanced Study}%
\centerline{Olden Lane, Princeton, N.J. 08540 USA}}}

\nref\one{A. Sagnotti, in {\it {Cargese '87}}, ``Nonperturbative
Quantum Field Theory'', eds. G.Mack {\it{et al.}} (Pergamon Press,
Oxford, 1988), p.521.}
\nref\two{P. Horava, Nucl. Phys. {\bf{B327}} (1989) 461\semi
Phys. Lett. {\bf {B231}} (1989) 251.}
\nref\three{M. Bianchi and A. Sagnotti, Phys. Lett. {\bf{B247}}
(1990) 517 \semi Nucl. Phys. {\bf{B361}} (1991) 519.}
\nref\four{E. Gimon and J. Polchinski, hep-th/9601038.}
\nref\five{A. Dabholkar and D. Park, hep-th/9602030.}
\nref\six {E. Gimon and C. Johnson, hep-th/9604129 \semi
A. Dabholkar and D. Park, hep-th/9604178.}
\nref\seven{M. Berkooz and R. Leigh, hep-th/9605049.}
\lref\eight{D. Morrison and C. Vafa, hep-th/9603161.}
\lref\nine{A. Sen, hep-th/9605150.}
\lref\ten{E. Gimon and C. Johnson, hep-th/9606176.}
\lref\eleven{M. Green and J. Schwarz, Phys. Lett. {\bf{B149}}
(1984)117.}
\lref\twelve{A. Sagnotti, Phys. Lett. {\bf{B294}} (1992)196.}
\lref\pol{J. Polchinski, hep-th/9606165.}
\lref\ferrara{S. Ferrara, J. Harvey, A. Strominger, and 
C. Vafa, Phys. Lett. {\bf{B361}}(1995)59\semi  
S. Ferrara, R. Minasian, and A. Sagnotti, hep-th/9604097.}
\lref\sunil{R. Gopakumar and S. Mukhi, private communication.}

\Title{\vbox{\baselineskip12pt
\hbox{IASSNS-HEP-96/74}\hbox{hep-th/9607019}}}
{\vbox{\centerline{An Orientifold from F Theory}}}

{\bigskip
\centerline{Julie D. Blum and Alberto Zaffaroni}
\bigskip
\ias

\bigskip
\medskip
\centerline{\bf Abstract}

The massless spectrum of an orientifold of the IIB string theory 
is computed and shown to be identical to F theory on the
Calabi-Yau threefold with $h_{11}=51$ and $h_{21}=3$.  Target
space duality is also considered in this model.}

\Date{6/96}

\newsec{Introduction}

Many recent and some older papers \refs{\one -
\seven} have explored 
orientifolds of the IIB string theory.  It is interesting to
expand the terrain of these orientifolds into the domains of
mother(male) and father(female) theories generally known
as M and F theories as we search for the ineluctably elusive
universal theory.

These orientifolds have the common feature that one divides
by a discrete symmetry group that includes world-sheet parity
and symmetries of space-time.  Cancelling tadpole anomalies
in these theories usually necessitates the addition of branes
but hopefully not branks.  In the case of orbifolds with 
larger than a $\bf {Z_2}$ space-time symmetry, one must be
careful of the definition of world-sheet parity in the
twisted sectors.  There are also subtleties in the application
of the IIB $SL(2,Z)$ (S) and target space (T) dualities.

We consider a simple orientifold here that illustrates the
relation to F theory and the subtlety of using T-duality.
The model we calculate here is F theory on the Calabi-Yau 
threefold with Hodge numbers $h_{11}=51$ and 
$h_{21}=3$ or $(51,3)$
which we show to be equivalent at the massless level to the 
$\bf{Z_2\times Z_2}$ orientifold of IIB $(\Omega (-1)^{F_l}R_3,
\Omega (-1)^{F_r}R_4)$ where $\Omega$ is world-sheet parity,
$(-1)^{F_l(F_r)}$ is left-handed(right-handed) space-time 
fermion number ($(-1)^{F_l}=e^{2\pi is_1^L}$), and $R_3(R_4)$
are space-time reflections on the torus $T^3(T^4)$ of the 
four-torus $T^3\times T^4 (R_3=e^{i\pi (s_3^L +s_3^R)},
R_4=e^{-i\pi (s_4^L +s_4^R)}$).  We have chosen $1,2$
as space-time directions and $3,4$ as internal
$T^4$ directions while $s$ denotes spin.
A future paper will look at the
$(3,51)$, which adds discrete torsion to the $(51,3)$;
S-duality; and other $\bf{Z_n}$ orientifolds.

\newsec{The Calculation}

First we consider the $(51,3)$ from the point of view of F theory.
As noted in \eight , this Calabi-Yau threefold can be obtained
{}from an orbifold of $T^6$ in which the two generators are $g_{35}$
and $g_{45}$ where $g_{ij}$ denotes reflections 
of $z_i$ and $z_j$ and $5$ is the
$11-12$ direction of F theory.  The results of Morrison and Vafa
imply that that the massless spectrum is $17$ tensors,
$4$ hypermultiplets, and a gauge group $SO(8)^8$.  

Now we wish to compute the IIB orientifold corresponding to 
F theory on the $(51,3)$. In a generic point of the 
moduli space this F theory
model is equivalent (by definition) to a compactification of 
type IIB on the
base of the elliptic fibration with a space-dependent 
complex coupling
constant identified with the complex structure of the fiber; 
seven-branes
are required at the singularities of the fibration. As shown by
Sen \nine , at the orbifold limit the coupling constant 
can be chosen to be
space-independent giving a standard type IIB compactification.  
A reflection
of $z_5$ is equivalent to 
a monodromy by $-1$ in $SL(2,Z)$.  By studying its action 
on the massless fields, we
recognize its equivalence to
the operator $\Omega (-1)^{F_l}$ of
the type IIB theory.

All the Voisin-Borcea models listed in \eight 
are the product of a two-torus
and $K_3$ divided by a $\bf{Z_2}$ symmetry; they correspond to type IIB
compactifications with a space-independent coupling constant. The
cases in which these Calabi-Yau are orbifolds can be exactly 
described by
type IIB orientifolds. The $\bf{Z_2\times Z_2}$ orbifolds 
are listed in \eight ,
and they correspond to $T^6$ divided by 
1) $g_{35}$ with a shift of order two
in $z_4$, $g_{45}$ with a shift of order two in $z_3$; 
2) the same except that
 $g_{45}$ is not accompanied by a shift; and 
3) the same with no shifts at all.
They have Hodge numbers $(11,11)$, $(19,19)$, and $(51,3)$ 
respectively.
The rule for obtaining the type IIB orientifold 
is quite simple: F theory
on the Voisin-Borcea model is the type IIB orientifold 
on $T^4$ obtained by
replacing in the above 
$\bf{Z_2\times Z_2}$ action the reflection of $z_5$ 
by $\Omega (-1)^{F_l}$ (or $\Omega (-1)^{F_r}$ 
if necessary to close the
algebra).

The case 1) is an F theory model with 9 tensors, 
 12 hypermultiplets, and no
gauge fields. By changing coordinates to $y_i=z_i+1/2, i=3,4$, the
orientifold is easily
seen to be equivalent to the following model: type IIB 
on $K_3$ divided by
 $\Omega (-1)^{F_l}\rho$ where $\rho$ is the Enrique's 
involution \ferrara .
This model makes sense as a closed string model. 
It is easy to check that
the Klein bottle tadpoles cancel. The shifts introduce an extra
$(-1)^m$ in the loop channel momentum sum
\eqn\shift{\sum_{m} (-1)^me^{-tm^2/R^2}}
which goes to zero in the limit $t\rightarrow 0$, as can 
be seen by a Poisson
resummation. There is no necessity to introduce open strings. 
By looking
at the action on the massless fields of $\Omega (-1)^{F_l}$ 
and at the action
of $\rho$ on the cohomology of $K_3$, we obtain exactly 9 tensors and
12 hypermultiplets.

The case 2) has $9$ tensors, $20$ hypermultiplets, 
and gauge group $U(1)^8$ at a generic point. 
{}From the 
orientifold point of view, one of the two operators involving $\Omega$
contains a space-time shift and does not produce tadpoles; 
the other one
requires $32$ seven-branes. As shown by R. Gopakumar and
S. Mukhi \sunil this is
the T-dual of the model in \five. 

The model 3) is the $\bf{Z_2\times Z_2}$
described in the introduction and is the more interesting since 
it has an 
unbroken gauge group even at a generic point of moduli space.
{}From the F theory point of view one gets an $SO(8)^8$ enhanced
gauge symmetry from the $D_4$ singularities of the fibration.
We want to understand this result from the orientifold point
of view.
This model is very similar to that of Gimon and Polchinski \four
except that two kinds of seven-branes are required rather than
five or nine-branes.  
As we will discuss extensively in the next section, 
this model cannot be 
obtained as the T- dual of the one in \four.

Let us first study the closed string spectrum. The calculation is 
straightforward, but we list the right-moving 
massless states since the
explicit form of the $\bf{Z_2}$ twisted sector will be useful in 
understanding
the subtleties of the model.
\eqn\spectri{\matrix{{\rm Sector}&{\rm State}&R&SO(4)
\,\,{\rm rep.}\cr\cr
{\rm NS}:&\psi^\mu_{-1/2}|0\!>&1&\bf{(2,2)}\cr
       &\psi^{1\pm}_{-1/2}|0\!>&-1&2\bf{(1,1)}\cr
       &\psi^{2\pm}_{-1/2}|0\!>&-1&2\bf{(1,1)}\cr
       {\rm R}:&|s_1s_2s_3s_4\!>&&&\cr
       &s_1=+s_2,\,s_3=+s_4&1&2\bf{(2,1)}\cr
       &s_1=-s_2,\,s_3=-s_4&-1&2\bf{(1,2)}\cr }}
       
and for the $\bf{Z_2}$ twisted sector:
\eqn\spectriii{\matrix{{\rm Sector}&{\rm State}&R&SO(4)
\,\,{\rm rep.}\cr\cr
{\rm NS}:&|s_3s_4\!>,\,\,s_3=+s_4&1&2\bf{(1,1)}\cr {\rm
R}:&|s_1s_2\!>,\,\,s_1=-s_2&1&\bf{(1,2)}.  }}

We have imposed the GSO projection and decomposed the little group of
the space-time Lorentz group as $SO(4)=SU(2){\times}SU(2)$. The
spectrum for the orientifold group is obtained 
by taking products of states from
the left and right sectors and dividing by $(\Omega (-1)^{F_l}R_3,
\Omega (-1)^{F_r}R_4, R)$.
The action of $R$ is listed in the table; $R_3=e^{i\pi (s_3^L +s_3^R)},
R_4=e^{-i\pi (s_4^L +s_4^R)}$, and
the $\Omega$ projection acts by symmetrizing 
left and right states in the Neveu-Schwarz--Neveu-Schwarz(NS-NS)
sector, while antisymmetrizing in the Ramond-Ramond(R-R) sector.

Let us note that the $\bf{Z_2\times Z_2}$ algebra 
closes only up to a factor
\eqn\ambig{(-1)^{F_l+F_r}\exp{2\pi i(s_i^L+s_i^R)}}
Fortunately, in the closed string Hilbert space this 
operator is identically $1$.
It acts as a global $-1$ in all the right or left sectors 
twisted by $1/2$,
but in the closed string Hilbert space it always cancels 
between left and right
states. However, this operator signalizes an ambiguity 
in the definition
of the $\bf{Z_2\times Z_2}$ algebra on the open string spectrum.

The closed string spectrum is now straighforward. 
{}From the untwisted sector we
have the supergravity multiplet, $1$ tensor and $4$ hypermultiplets. 
The main difference between this model and that of \four 
(clearly showing that the two models 
are not T-dual) lies in the twisted sector 
where $(-1)^{F_l}R_3$ introduces
an extra minus sign; as a consequence, now
 the hypermultiplets are projected out,
and we get $16$ tensors, one from each fixed point.

Let us now turn to the open string spectrum. 
The tadpoles are essentially the same; the presence of $(-1)^{F_l}$ 
compensating in the Ramond sector for 
extra signs introduced by $R_{3,4}$.  Therefore, we omit 
the calculation and present only the results.
First, we list the tadpoles for the untwisted R-R  potentials. We have
 (proportional to
$(1-1){v_{3}\over 16v_4}\int_0^\infty dl$):

\eqn\tadsone{\Tr(\gamma^{\phantom{-1}}_{0,7})^2 - 
64\Tr(\gamma_{\Omega R_3(-1)^{F_l},7}^{-1}
\gamma_{\Omega R_3(-1)^{F_l},7}^T) + 32^2,}

(proportional to 
$(1-1){v_4\over 16v_3}\int_0^\infty dl$):

\eqn\tadstwo{\Tr(\gamma^{\phantom{-1}}_{0,7'})^2 - 64
\Tr(\gamma_{\Omega R_3(-1)^{F_l},7'}^{-1}
\gamma_{\Omega R_3(-1)^{F_l},7'}^T) + 32^2,}

and for the twisted potentials

\eqn\ttad{{1\over 16}\sum_{IJ}(\Tr\gamma_{R7,I} - \Tr\gamma_{R7',J})^2}

where $I$ and $J$ are respectively the four fixed points 
of $R_3$, the four
of $R_4$, and other notations are those of \four .

The number of seven-branes is $32$ for each of the two kinds.  There
are a couple of subtleties in the calculation which caused us a bit of
trouble.  Using the same argument as Gimon and Polchinski for the sign
of $\Omega^2$ in the $5-9$ sector, we obtain an extra minus sign for
$\Omega^2$ acting on the seven-brane vacuum.  This extra minus is
confirmed by the new consistency condition discovered in \pol . This
condition reads (for the model of ref. \four ) \eqn\cond{\gamma_R = -
\gamma_{\Omega}\gamma_R^T\gamma_{\Omega}^{-1},} and it is obtained by
considering the transition between the closed string R-R twisted
ground state and the open string vector.  In our case $\Omega$ is
replaced by $\Omega(-1)^{F_l}R_3$ and $(-1)^{F_l}$ introduces an extra
minus sign; therefore $\gamma_R$ can be chosen to be symmetric.  We
must be careful with the definition of the operators in the open
string sector since, as we noted before, there is an ambiguity in the
closure of the $\bf{Z_2\times Z_2}$ algebra (\ambig).  This ambiguity
is only revealed in the $7-7'$ sector.  Multiplying together the two
$\bf{Z_2}$ operators involving $\Omega$ gives $\Omega^2 R$ rather than
$R$ for open string states in the $7-7'$ sector.  This $\Omega^2 R$
does not affect the tadpole equations where $R$ is used since we want
to cancel divergences of the NS-NS and R-R sectors of the closed
string, but gives an extra minus in the $7-7'$ sector.  The algebraic,
tadpole equations, and the consistency equation discovered in \pol are
satisfied while the $\bf{Z_2\times Z_2}$ algebra is realized if we
choose the matrices operating on Chan-Paton factors equal to the
identity.  The twisted tadpole condition \ttad is satisfied if we put
eight seven-branes at each of the fixed points of $R_3$ and $R_4$.
All the matter is projected out.  So the open string sector yields
simply the gauge group $SO(8)^8$.  As we discussed before, we also get
one tensor and four hypermultiplets from the untwisted closed string
sector as well as sixteen tensors from each of the fixed points of $R$
in the $R$-twisted sector.  Thus, the orientifold agrees with the
result of F theory.

\newsec{T-Duality}

We noted before that we cannot map this model 
to a simpler one by a T-duality.
In fact, let us
 apply T-duality in the three direction.  The effect
is to transform an operator $O$ to 
$e^{i\pi s_3^L} O e^{-i\pi s_3^L}$ so that operators 
with $\Omega$ get an extra factor $e^{i\pi
(s_3^L -s_3^R)}$.  Other operators are unchanged.  It is
easy to show that the spectrum remains unchanged under the 
transformation.  

Notice that $\Omega$ is transformed to
$\Omega R_3 e^{2\pi is_3^L}$. The
Gimon and Polchinski \four model has a T-dual model which contains two
kinds of seven-branes but differs from the $(51,3)$ model.  
The difference is that in the definition of the projection operators
every occurrence of $e^{2\pi is_3^L}$ is replaced by $(-1)^{F_l}=
e^{2\pi is_1^L}$. These two operators are equivalent 
whenever there are no
twisted sectors, but they differ by a minus sign in every $Z_2$ twisted
sector as is obvious from equation \spectriii .
We have already utilized this observation when we computed 
the twisted
contributions to the closed string spectrum. 
Gimon and Polchinski had one
hypermultiplet from each fixed point, while we got a tensor. 
Thus, we obtain a different
answer from \ten because of the T-duality subtlety.  The 
irreducible part of the gravitational anomaly is cancelled
by the spectrum we have obtained.  There is no Higgs mechanism
as there is no charged matter.  The sixteen tensors obtained 
in the $R$-twisted sector should suffice to cancel any left
over anomalies by the methods of \eleven\twelve.  This model
has illustrated the application of Gimon and Polchinski
techniques in F theory and serves as a starting point
for understanding other models.

\bigskip\centerline{\bf Acknowledgements}\nobreak

We thank K. Dienes, M. Flohr, R. Gopakumar, A. Hananay,
K. Intriligator, C. Johnson, S. Mukhi,
 A. Sen, and E. Witten for
useful discussions.  This research was supported in part by
NSF Grant PHY-9513835 and DOE DE-FG02-90ER40542.

\listrefs
\end